\documentclass[aps,prd,twocolumn,groupedaddress,preprintnumbers,superscriptaddress,nofootinbib]{revtex4}
\usepackage{amsmath,amssymb,latexsym}

\begin{document}

\title{
Conformal Inflation, Modulated Reheating, and WMAP5
}


\author{
Takeshi Kobayashi
}\email[]{tkobayashi@utap.phys.s.u-tokyo.ac.jp}
\affiliation{
Department of Physics, School of Science, The University of Tokyo,
Hongo 7-3-1, Bunkyo-ku, Tokyo 113-0033, Japan
}
\author{
Shinji Mukohyama
}\email[]{shinji.mukohyama@ipmu.jp}
\affiliation{
Institute for the Physics and Mathematics of the Universe (IPMU), The
University of Tokyo, 5-1-5 Kashiwanoha, Kashiwa, Chiba 277-8582, Japan
}

\preprint{UTAP-605} 
\preprint{RESCEU-59/08}
\preprint{IPMU 08-0060}


\begin{abstract}
 We investigate density perturbations generated through modulated
 reheating while inflation is driven by a conformally coupled scalar
 field. A large running of the spectral index is obtained,
 which reflects the basic nature of conformal inflation that
 higher-order time derivatives of the Hubble parameter during inflation
 are not necessarily small. This feature may allow us to distinguish
 between conformal inflation models and standard minimally coupled
 ones. We also investigate how the resulting fluctuations are modified
 when there is a deviation from an exact conformal coupling between the
 inflaton and gravity. Finally, we apply our results to the warped
 brane inflation model and see that observational bounds from the WMAP5
 data suggest a blue tilted density perturbation spectrum. 
\end{abstract}

\maketitle

\section{Introduction}
\label{sec:intro}

Ever since the idea of cosmic inflation was proposed, 
inflationary model building has largely focused on slow-rolling scalar
fields minimally coupled to gravity as candidate inflatons. However,
recently it was pointed out in \cite{Kofman:2007tr} that conformally
coupled scalar fields are also capable of accelerating the universe. 
Since the existence of such conformally coupled fields is rather common
in models from string theory, it is of great interest to explore the
possibility of conformally coupled fields driving inflation. In this
light, we also come up with a new question of how we can distinguish
between these ``conformal inflation'' models and standard minimally
coupled ones.   

The aim of this paper is to focus on density perturbations and seek
distinctive features conformal inflation might have left. A
conformally coupled inflaton itself cannot be responsible for generating
primordial fluctuations\footnote{From a conformally coupled inflaton,
one generally obtains a highly blue tilted spectrum. (This can easily be
seen by moving to the Einstein frame and computing the density
perturbations~\cite{Makino:1991sg}.) Therefore, unless we consider
inflation with an extremely high energy scale, perturbations generated by
the inflaton will be suppressed at the CMB scale.}, but
instead string theory suggests alternative scenarios such as modulated
reheating~\cite{Dvali:2003em,Kofman:2003nx}. Hence, in this paper we
consider the case where modulated
reheating generates density perturbations while inflation is driven by
an almost conformally coupled inflaton.\footnote{One may wonder whether
the coupling of the inflaton to Standard Model particles may spoil
conformal symmetry. However, we need not require an exact symmetry,
instead, the basic factors for conformal inflation are the inflaton's
(almost) conformal coupling to gravity together with an appropriate
inflaton potential.} We study the scale dependence of
the generated density perturbations. The general results we obtain are
that (i) modulated reheating together with conformal inflation can
produce a nearly scale-invariant spectrum, and (ii) the running of the
spectral index~$|dn_s/d \ln k|$ turns out to be as large as $|n_s-1|$. The
latter result reflects the nonexistence of hierarchy among 
higher-order time derivatives of the Hubble parameter $d^n\ln H^2 / dt^n
H^n $ during conformal inflation. This is in 
strong contrast to standard minimal models where higher-order
derivatives are suppressed by higher orders of the slow-roll parameters
(and their derivatives). Hence this feature offers a chance to obtain
a smoking-gun signal for nonminimally coupled inflation models.

We also investigate how the resulting density perturbations are modified
when there is a deviation from an exact conformal coupling between the
inflaton and gravity. The result will allow us to impose constraints on
the inflaton's coupling from observational data. 

As a concrete example, our generic results are applied to the
well-studied warped brane inflation
model~\cite{Kachru:2003sx,Baumann:2006th,Baumann:2007np,Krause:2007jk,Baumann:2007ah,Baumann:2008kq}. We
will see that comparison with the WMAP5 data~\cite{Komatsu:2008hk}
predicts a blue tilt ($n_s >1$) for this model.

The paper is organized as follows. First we give a quick review on
conformal inflation in Section~\ref{sec:revcon}. In
Section~\ref{sec:dpmr}, we study the scale dependence of the density
perturbations generated through modulated reheating after conformal
inflation. Then in Section~\ref{sec:wbi} we apply our results to the
warped brane inflation model. We conclude in Section~\ref{sec:conc}. In
the appendix we provide detailed arguments on conformal
inflation.

\section{Review of Conformal Inflation}
\label{sec:revcon}

Here we give a brief review of conformal inflation. Consider the action
\begin{equation}
 S=\int dx^4\sqrt{-g} \left[
		       \frac{M_p^2}{2}\mathcal{R}-\frac{1}{2}g^{\mu\nu}
		       \partial_{\mu} \phi \partial_{\nu} \phi - 
			V(\phi) -\frac{\xi}{2} \mathcal{R}
			\phi^2\right] \label{action}
\end{equation}
where $\mathcal{R}$ is the scalar curvature and $\xi$ is the nonminimal
coupling to gravity. Choosing a flat FRW background,
\begin{equation}
 ds^2 = -dt^2 + a(t)^2 d\mathbf{x}^2,
\end{equation}
the Friedmann equation is
\begin{equation}
 ( M_p^2 - \xi \phi^2 ) H^2
= \frac{1}{6} \dot{\phi}^2 + \frac{1}{3} V(\phi) + 2 \xi H 
\phi \dot{\phi}, \label{1}
\end{equation}
and the equation of motion of $\phi$ is
\begin{equation}
 \ddot{\phi}+ 3 H \dot{\phi} + V'(\phi) + 6\xi (\dot{H}+2 H^2)\phi
  =0. \label{3}
\end{equation}
By introducing 
\begin{equation}
 \pi \equiv \dot{\phi} + H\phi, 
\end{equation}
the equations (\ref{1}) and (\ref{3}) can be rewritten in the following
form:
\begin{equation}
 M_p^2 H^2 = \frac{1}{3} V + \frac{1}{6} \pi^2 + \left(\xi -
						\frac{1}{6}\right) (2 H
 \phi \pi -   H^2\phi^2) , \label{11}
\end{equation}
\begin{equation}
 \dot{\pi}+ 2 H \pi + V' + 6 \left(\xi - \frac{1}{6}\right) (\dot{H}+2
  H^2) \phi = 0. \label{33} 
\end{equation}

The conformal case $\xi = 1/6$ was investigated in
\cite{Kofman:2007tr}, where inflation was realized while the equations
(\ref{11}) and (\ref{33}) could be approximated to
\begin{equation}
 M_p^2 H^2 \simeq \frac{1}{3} V, \label{111}
\end{equation}
\begin{equation}
 c H \pi \simeq -V'. \label{333}
\end{equation}
Here, $c$ is a dimensionless constant whose meaning will soon become 
clear\footnote{Our parameterization of $c$ differs from that of
\cite{Kofman:2007tr} by $c_{\mathrm{ours}} =
c_{\mathrm{[1]}}+2$.}. Let us define three ``flatness
parameters'' as
\begin{align}
 \epsilon &\equiv \frac{M_p^2}{2}\left(\frac{V'}{V}\right)^2,
 \label{epsilon} \\ 
 \tilde{\epsilon} &\equiv \frac{\phi V'}{2 V}, 
 \label{tepsilon} \\
 \eta_c &\equiv \eta + \frac{c}{3}\left(\frac{V''\phi}{V'}+c-2\right),
 \label{etac}
\end{align}
where
\begin{equation}
 \eta \equiv M_p^2 \frac{V''}{V}.
\end{equation}
Then one can check that the necessary conditions for the
approximations (\ref{111}) and (\ref{333}) to be valid are 
\begin{equation}
 \frac{6}{c^2} \epsilon\ll 1,\quad
\left| \frac{6}{c^3}\epsilon + \frac{1}{c}\tilde{\epsilon} -
  \frac{3}{c^2} \eta_c \right| \ll 1.
\end{equation}
The necessary condition for $|\dot{H}/H^2| \ll 1$ can also be
derived,
\begin{equation}
 \left| \frac{6}{c^2}\epsilon + \tilde{\epsilon}\right| \ll 1.
\end{equation}
When $|c| \sim \mathcal{O}(1)$, these conditions take the simple form
\begin{equation}
 \epsilon \ll 1,\quad  |\tilde{\epsilon}| \ll 1, \quad |\eta_c | \ll 1.
 \label{ci-cond}
\end{equation}
The proportionality constant~$c$ is chosen such that it is the largest
constant to minimize $|\eta_c|/c^2$. For detailed arguments, see
Appendix~\ref{app:A1}.

It is clear from (\ref{11}) and (\ref{111}) that $\pi^2/V$ is small
during inflation. Especially when the potential is flat enough to
satisfy $|M_p^2 V'| \ll |c \phi V|$, then
\begin{equation}
 \dot{\phi} \approx -H \phi,
\end{equation}
which suggests that the inflaton is rapidly rolling towards its origin.

\section{Density Perturbations from Modulated Reheating}
\label{sec:dpmr}

We analyze the case where density perturbations are generated through
modulated reheating~\cite{Dvali:2003em,Kofman:2003nx} after conformal
inflation. The scale dependence of the perturbations will be expressed in
terms of the flatness parameters and $(\xi - 1/6)$. 

\subsection{Modulated Reheating Scenario}
\label{subsec:mrs}

To reheat the universe, the inflaton should decay into ordinary Standard
Model particles (or particles that eventually transfer their energy into
SM particles) after inflation. Here, it may well be that the decay  
rate of the inflaton field (i.e. couplings between the inflaton and
ordinary particles) is determined by VEVs of fields
in the theory. If those fields are light during inflation, they will
fluctuate and generate spacial fluctuations in the decay rate for the
reheating process, which translate into density perturbations in the
universe. 

Let us take $\chi$ to be the light modulus field whose VEV determines the
coupling~$\lambda$. Then its fluctuations during inflation manifest
themselves as fluctuations in $\lambda$ and the decay rate~$\Gamma$, leading
to density perturbations
\begin{equation}
 \frac{\delta\rho}{\rho} \propto \frac{\delta \Gamma}{\Gamma}
 \propto \frac{\delta \lambda}{\lambda} \propto \frac{\delta\chi}{\chi}.
\end{equation}
Fields whose mass is smaller than $H$ during inflation typically
obtain fluctuations of order $H$. Therefore the tilt of the spectrum in
this scenario can be expressed in a form independent of the details of
the model,
\begin{equation}
 n_s - 1 = \frac{d\ln H^2}{d \ln k}= 2
  \frac{\dot{H}}{H^2} \left(1+\frac{\dot{H}}{H^2}\right)^{-1},
  \label{ns} 
\end{equation}
\begin{equation}
 \frac{d n_s}{d \ln k} = 
2\left(1+\frac{\dot{H}}{H^2}\right)^{-3}\frac{1}{H}
\left( \frac{\dot{H}}{H^2}\right)^{\cdot},
 \label{running}
\end{equation}
where the spectral index and its running are estimated at the moment of
horizon crossing~$k = aH$. 

We do not restrict ourselves to specific setups of modulated reheating
in this paper. Rather, we focus on the generic features~(\ref{ns}),
(\ref{running}) of modulated reheating and study the imprints of
conformal inflation on the tilt of the spectrum. (Some candidates for
the light modulus field in the case of warped brane inflation are
proposed in Section~\ref{sec:conc}.)

\subsection{The Spectral Index and its Running}
\label{subsec:sir}

In this subsection, the observables are expanded in terms of the three
flatness parameters $\epsilon$, $\tilde{\epsilon}$, and $\eta_c$. We
assume $|c|\sim \mathcal{O}(1)$. From now on, we take
account of the inflaton's slight deviation from a conformal coupling~$(
\xi -1/6)$. This procedure allows us to deal with a wide variety of 
situations, e.g., when there exists additional corrections to the action
which ruin the exact conformal coupling, when the frame where the
inflaton~$\phi$ is conformally coupled to gravity differs from the frame
where the light modulus~$\chi$ is minimally coupled. 

Since the flatness parameters only describe the form of the potential, let
us introduce an additional dimensionless parameter~$\alpha$ as follows,
\begin{equation}
 -V' = c H \pi + \frac{V}{M_p}\alpha. \label{alpha}
\end{equation}
First we make use of this parameter to evaluate quantities, and then later
on we will estimate the value of $\alpha$ itself. For convenience, we
further define
\begin{equation}
 e \equiv \frac{M_p V'}{\sqrt{2} V},\quad  \kappa \equiv \frac{\phi
  }{M_p}, \quad \sigma \equiv \frac{\pi^2}{V}. 
\end{equation}
Note that $\epsilon\!=\!e^2$ and $\tilde{\epsilon}\! =\!
e\kappa/\sqrt{2}$. Then (\ref{alpha}) can be rewritten as 
\begin{equation}
 \sigma = \frac{V}{c^2 M_p^2 H^2}\left( \alpha + \sqrt{2} e\right)^2.
 \label{b}
\end{equation}
Also, from (\ref{11}) and (\ref{alpha}) one can derive
\begin{equation}
 \frac{3 M_p^2 H^2}{V} = 
 \frac{1+\frac{1}{2}\sigma - \frac{ 1}{c}(6\xi-1)
  \kappa \left( \alpha +\sqrt{2}e\right)}{
1+\frac{1}{6}\left(6\xi-1\right)\kappa^2}.
\label{a}
\end{equation}
(\ref{b}) and (\ref{a}) give a quadratic equation on $\sigma$,
whose solution is
\begin{equation}
 \sigma =-1+X+(6 \xi-1)\frac{\kappa (\alpha +\sqrt{2}e)}{c} ,
\end{equation}
where
\begin{widetext}
\begin{equation}
 X \equiv \sqrt{\frac{
 \left\{c-(6\xi-1) \kappa(\alpha +\sqrt{2}e)\right\}^2
+  (\alpha+ \sqrt{2}e)^2 \left\{6+ (6\xi-1) \kappa^2\right\}
}{c^2}}
\end{equation}
is a quantity close to one. We have chosen the
solution for $\sigma$ to approach zero in the $\epsilon,
\tilde{\epsilon}, \alpha, (\xi\!-\!1/6) \to 0$ limit. The time
derivative of the Hubble parameter is given by (\ref{11}) and (\ref{33})
\begin{equation}
 \frac{\dot{H}}{H^2} =- \frac{2\sigma + 12 \xi \tilde{\epsilon} + (6 \xi
  - 1) \left\{ -\sigma + 6 \frac{ H \phi \pi}{V} + 3 (4 \xi-1) \frac{H^2
	\phi^2}{V}\right\}} 
{2+\sigma + (6 \xi-1) \left\{ 2 \frac{H \phi \pi}{V} + (6 \xi-1)
		       \frac{H^2 \phi^2}{V}\right\}}.
\label{dotH}
\end{equation}
Using the results above, (\ref{dotH}) can be evaluated in terms of $e, 
 \tilde{\epsilon},  \kappa, \xi$ and $\alpha$,
\begin{equation}
= \frac{
 2c (1-X-\tilde{\epsilon})+(6\xi-1)
 \left[3\kappa (\alpha +\sqrt{2}e)
 \left\{1+\xi \left(2+(6\xi-1) \kappa^2\right)\right\}
 -c\left\{1-X+2 \tilde{\epsilon} -\kappa^2 +\kappa^2
    \xi(3+X+2\tilde{\epsilon})\right\} 
\right]
}{
\left\{
 c(1+X)-(6\xi-1)\kappa(\alpha +\sqrt{2}e)
\right\}
 \left\{ 1+ \xi (6\xi-1)\kappa^2
\right\}
}. \label{dotH-alpha}
\end{equation}
Here it should be noted that $\alpha$ shows up either in the quadratic
 form~$\alpha^2$, or with a prefactor $e\alpha$, $(6\xi-1)\alpha$. The
 second order derivative is given by 
\begin{multline}
 \frac{1}{H}\left(\frac{\dot{H}}{H^2}\right)^{\cdot} = 
 - \frac{6
}{
2+\sigma + (6\xi-1)\left\{ 2 \frac{H\phi \pi}{V} + (6\xi-1) \frac{H^2
		    \phi^2}{V}\right\} 
}
\Biggl[
(10\xi-3)\sigma -2 (7\xi-1)
 \tilde{\epsilon}+ (3\xi-1) \frac{\pi V'}{HV}
+\xi\eta \left(\frac{\phi \pi}{M_p^2 H}-\kappa^2 \right) \\
 +(6\xi-1) \left\{ 2 (4\xi-3) \frac{H\phi\pi}{V}-(16\xi-3)
	    \frac{H^2\phi^2}{V} \right\} +
 \frac{\dot{H}}{H^2} \Biggl\{ (2\xi-1)\sigma -4 \xi \tilde{\epsilon} 
 + (6\xi-1) \left\{ 2 (2\xi-1) \frac{H\phi \pi}{V} - (8\xi-1)
	     \frac{H^2\phi^2}{V}\right\} \Biggr\}
\Biggr].
\label{ddotH}
\end{multline}
\end{widetext}
To avoid clutter, we do not lay out the explicit form of this
quantity. However, one can check that $\alpha$ enters (\ref{ddotH}) 
through $\alpha^2$, $e\alpha$, and $(6\xi-1)\alpha$. 

Now we turn to evaluating $\alpha$ itself. Though it is
difficult to obtain the explicit value of $\alpha$, we can still derive
its evolution 
equation. By differentiating both sides of (\ref{alpha})
with respect to time and then substituting (\ref{33}), one obtains
\begin{multline}
 \frac{\dot{\alpha}}{H} = 
 \alpha \left\{ -2+2 \tilde{\epsilon} +\frac{\dot{H}}{H^2}
+ (\eta-2 \epsilon) \frac{V}{c M_p^2 H^2} \right\} \\
+\sqrt{2} e \left\{ c \sigma + \frac{\dot{H}}{H^2}
 +\frac{3}{c}\left( \eta_c - 2\epsilon +\eta
 \frac{V-3M_p^2 H^2}{3 M_p^2 H^2}\right)  \right\}  \\
+ 2\kappa \left(\xi - \frac{1}{6}\right) \left(2+\frac{\dot{H}}{H^2}
\right) \frac{3 M_p^2 H^2}{V}, \label{alpha-diff}
\end{multline}
where we have made use of 
\begin{equation}
 \eta \kappa = \sqrt{2} e \left\{
\frac{3}{c} (\eta_c - \eta)+2-c\right\} 
 \label{eta-kappa}
\end{equation}
along the way. The results above allow us to evaluate the r.h.s. of
(\ref{alpha-diff}) in terms of $e, \epsilon, \tilde{\epsilon},
\eta_c, \eta, \kappa, \xi$ and $\alpha$. First let us consider the limit
where exact conformal coupling is obtained and the three flatness
parameters are negligible. Then the evolution equation~(\ref{alpha-diff})
turns into 
\begin{align}
 \frac{\dot{\alpha}}{H} &= \alpha\frac{6 c\eta-4\sqrt{c^4+6 
  c^2   \alpha^2}}{c^2+\sqrt{c^4+6 c^2\alpha^2}} 
 \label{18}\\
 &\sim \alpha \left(-2+\frac{3\eta}{c}\right), 
\end{align}
where the last line is an approximation valid in the case where
$|\alpha |\ll |c| $. This shows that when $-2+3\eta/c < 0$, $\alpha$
damps as the universe expands during the inflationary era. When $\alpha$
becomes as small as the terms we have neglected in obtaining~(\ref{18}),
the value of $\alpha$ settles down to that corresponding to the
neglected source terms. Expanding the r.h.s. of (\ref{alpha-diff}) in
terms of the flatness parameters and $(\xi\!-\!1/6)$, one can show
\begin{multline}
 \frac{\dot{\alpha}}{H} = \mathcal{O}(\epsilon,\, \epsilon^{1/2} \eta_c)
 + \mathcal{O}\left(\xi- \frac{1}{6}\right) \\
\quad  + \alpha \left\{-2+\frac{3\eta}{c}+\mathcal{O}(\epsilon^{1/2})+
	    \mathcal{O}\left(\xi-\frac{1}{6}\right) \right\} +
	    \mathcal{O}(\alpha^2).   \label{25}
\end{multline}
Here, contributions from a nonexact conformal coupling are denoted by
$\mathcal{O}((\xi\!-\!1/6)^n)$, which refers to products of 
$\mathcal{O}(1)$ factors and $(\xi\!-\!1/6)^m$ with $m\ge n$. It should
be noted that we have not imposed any assumptions on the values of
$\kappa$ and $\eta$. (\ref{25}) shows that during inflation the value of
$\alpha$ becomes of the order
\begin{equation}
 \alpha =
 \mathcal{O}(\epsilon,\, \epsilon^{1/2} \eta_c)
 + \mathcal{O}\left(\xi- \frac{1}{6}\right).
\label{alpha-order}
\end{equation}

Now that we know the value of $\alpha$, we can compute the spectral
index~(\ref{ns}) and its running~(\ref{running}) in the modulated
reheating scenario. Recalling how $\alpha$ showed up in time derivatives
of the Hubble parameter, the leading order behavior of the tilt can be
described without $\alpha$. We obtain
\begin{widetext}
\begin{equation}
 n_s -1 = 
 -2\tilde{\epsilon} - \left( \frac{12}{c^2} + \kappa^2\right) \epsilon +
 \mathcal{O}(\epsilon^{3/2},\, \epsilon \eta_c) +
 \left(\xi-\frac{1}{6}\right)  
 \left\{ 2\kappa^2 + \mathcal{O}(\epsilon^{1/2})\right\} 
 + \mathcal{O}\left(\left( \xi - \frac{1}{6}\right)^2\right) ,
 \label{result1}
\end{equation}
\begin{multline}
 \qquad 
 \frac{d n_s}{d \ln k}=
 2\left(3-c+\frac{3}{c} \eta_c \right)\tilde{\epsilon} + \left( \frac{6
   (8-3c)}{c^2} + (7-3c) \kappa^2 
 \right)\epsilon + \mathcal{O}(\epsilon^{3/2},\, \epsilon \eta_c) \\
 + \left(\xi - \frac{1}{6}\right) \left\{-4\kappa^2 +
 \mathcal{O}(\epsilon^{1/2})\right\} 
 +\mathcal{O}\left(\left( \xi -
 \frac{1}{6}\right)^2\right). 
 \qquad 
 \label{result2} 
\end{multline}
\end{widetext}
We immediately see that $|dn_s /d\ln k|$ can become large,
comparable to $|n_s-1|$. This is 
due to the fact that the derivatives of the flatness parameters do not
necessarily become smaller than the parameters themselves. Since this is
a generic feature of conformal inflation, a large running is expected
even if we consider mechanisms other than modulated reheating for
generating fluctuations (e.g. curvaton
models~\cite{Enqvist:2001zp,Lyth:2001nq,Moroi:2001ct}). Also, our
results indicate 
that $(\xi\!-\!1/6)$-corrections can dominantly determine the values of the
cosmological observables unless $|\xi\!-\!1/6 |$ is smaller than the
flatness parameters. 

Before ending this section, we should remark that when the inflaton is
extremely close to its origin (i.e. $\kappa^2 \lesssim
\mathcal{O}(\epsilon^{1/2})$), one may need to solve the evolution 
equation~(\ref{alpha-diff}) of $\alpha$ in order to compute the
$(\xi\! -\! 1/6)$-corrections to the observables, since the ``subleading''
terms denoted by $(\xi\! -\! 1/6)\mathcal{O}(\epsilon^{1/2})$ in
(\ref{result1}) and (\ref{result2}) become important. For example,
(\ref{eta-kappa}) shows that such situations are realized
when $|\eta | \gtrsim 1$ or when $c\approx 2$. See also
Appendix~\ref{app:A2} for arguments on the validity of our results
before the inflaton trajectory approaches the attractor.

\section{Application to Warped Brane Inflation}
\label{sec:wbi}

Let us now apply the results obtained in the previous section to a
specific model. Here we consider the warped brane inflation
model~\cite{Kachru:2003sx,Baumann:2006th,Baumann:2007np,Krause:2007jk,Baumann:2007ah,Baumann:2008kq},
where the universe experiences inflation while a D3-brane moves towards
the tip of a flux compactified warped throat. The D3-brane is pulled by
a stack of 
$\overline{\mathrm{D}3}$-branes sitting at the tip. If the position of
the D3-brane is a conformally coupled scalar (for a discussion on this
issue, see e.g. \cite{Seiberg:1999xz}), then this model serves as a
realization of conformal inflation. The attractor behavior in this case
is demonstrated in \cite{Kofman:2007tr} by the phase portrait method. 

Considering a throat whose geometry is $AdS_5 \times X_5$, the potential
of the inflaton takes the form 
\begin{equation}
 V(\phi) = 2 p h_0^4 T_3\left(1-\frac{h_0^4 T_3^2 R^4}{N \phi^4}\right).
\end{equation}
Here, the inflaton is related to the radial position~$\rho$ of
the D3 through $\phi\! =\! \sqrt{T_3} \rho$, $p$ is the number of
$\overline{\mathrm{D}3}$s at the tip, $h_0\! =\! \rho_0/R$ is the
warping at the tip, $T_3\! =\! 1/(2\pi)^3 g_s (\alpha ')^2$ is the D3
tension, $R^4\!=\! 2^2 \pi^4 g_s (\alpha')^2 N/ \mathrm{Vol}(X_5)$ is the
AdS radius of the throat, $\mathrm{Vol}(X_5)$ is the dimensionless
volume of the base space~$X_5$, and $N (>1)$ is the 5-form charge. 

The conformally coupled inflaton satisfies $\dot{\phi} \approx -\phi H$
during inflation, hence the number of $e$-foldings generated
is\footnote{The deviation from an exact conformal coupling will appear
as an $\mathcal{O} (\xi\!-\!1/6)$ correction to this result, which we can
ignore.} 
\begin{equation}
 \mathcal{N} \approx \log \left(\frac{\rho_i}{\rho_f}\right) \approx -
  \log\left( \frac{h_0}{\lambda_i }\right),
\end{equation}
where we have set the initial position of the D3 by a
dimensionless constant~$\lambda_i$ as $\rho_i = \lambda_i R$, and
assumed inflation to end when the D3 approaches the tip $\rho_f 
\approx \rho_0$. This shows that in order to obtain enough $e$-foldings,
the throat should be strongly warped $h_0/\lambda_i \ll 1$.

Assuming $\alpha$ in (\ref{alpha}) to be sufficiently damped when the
perturbations of the CMB scale are generated, the scale dependence of the
perturbations can be computed by (\ref{result1}) and (\ref{result2}). We
parametrize the position of the D3 when the CMB scale was originally
produced by $\rho_{\mathrm{CMB}} = \lambda R$. Furthermore,
for simplicity, we ignore the compactified bulk to which the throat is
glued, hence
\begin{equation}
 M_p^2 \simeq \frac{2}{(2\pi)^7g_s^2(\alpha ')^4} \int^{R}_{\rho_0} d\rho \,
 \mathrm{Vol}(X_5) \frac{\rho^5}{h^4} \simeq \frac{\mathrm{Vol}(X_5)
 R^6}{(2\pi)^7 g_s^2 (\alpha')^4}.
\end{equation}
Then under the assumption $h_0/\lambda \ll 1$, the flatness parameters
and $\kappa$ can be calculated (note that $c=7$),
\begin{equation}
 \epsilon \simeq \frac{2 h_0^8}{\lambda^{10} N},\
\tilde{\epsilon}  \simeq \frac{2h_0^4}{\lambda^4 N},\
\eta_c = \eta \simeq -\frac{5h_0^4}{\lambda^6}, \
\kappa^2 \simeq \frac{4\lambda^2}{N}.
\end{equation}
Since $\epsilon$ is extremely small compared to the other flatness
parameters, the cosmological observables can be estimated as follows:
\begin{align}
 n_s -1 &\simeq -2 \tilde{\epsilon} + 2  \kappa^2  \left( \xi -
 \frac{1}{6}\right) \\
&\simeq -\frac{4}{N}  \left\{  \frac{h_0^4}{\lambda^4}-2 \lambda^2
				 \left( \xi - \frac{1}{6}\right)\right\}  
\end{align}
\begin{align}
 \frac{d n_s}{d \ln k} &\simeq  -8\tilde{\epsilon} - 4
  \kappa^2   \left( \xi - \frac{1}{6}\right) \\
 &\simeq -\frac{16}{N} \left\{ \frac{h_0^4}{\lambda^4}+
   \lambda^2\left( \xi - \frac{1}{6}\right) \right\} .
\end{align}

The 5-year WMAP+BAO+SN data give bounds $n_s = 1.022_{-0.042}^{+0.043}$
(68\% CL) and $dn_s/d\ln k = -0.032_{-0.020}^{+0.021}$ (68\% CL) when
tensor mode perturbations are negligible~\cite{Komatsu:2008hk}. 
Since the $h_0^4/\lambda^4$ terms are too small to be constrained by the
observational bounds, we ignore them. Then the observables are
determined only by the inflaton's deviation from an exact conformal
coupling, 
\begin{equation}
 n_s-1 \simeq 8 \frac{\lambda^2}{N} \left(\xi - \frac{1}{6}\right),\ 
 \frac{d n_s}{d \ln k} \simeq -16 \frac{\lambda^2 }{N} 
 \left(\xi -  \frac{1}{6}\right).
\end{equation}
It is easy to see that the spectral index and its running are related by
\begin{equation}
  \frac{d n_s}{d \ln k} \simeq -2 (n_s-1). \label{relation}
\end{equation}

Using the observational bound on $dn_s/d\ln k$ to constrain  
$\lambda^2(\xi\!-\!1/6)/N$, 
\begin{equation}
 0.001 \lesssim \frac{\lambda^2}{N} \left( \xi - \frac{1}{6}\right)
  \lesssim   0.003 . \label{xi-bound}
\end{equation}
If we further make use of this bound to constrain the spectral index, we
obtain 
\begin{equation}
 0.006 \lesssim n_s-1 \lesssim 0.026.
\end{equation}
Thus the 1$\sigma$ observational bound on the running allows us to
predict a blue tilt for the warped brane inflation model. Moreover, the
values of the observables were dominantly determined by the inflaton's
deviation from a conformal coupling. The bound~($\ref{xi-bound}$)
suggests that this deviation can be fairly large, e.g., $(\xi\!-\!1/6) \sim
10^{-1}$ when $\lambda \sim 1$, $N \sim 10^2$. However, note that
if $N$ ($\lambda$) is larger (smaller), then such a large
$(\xi\!-\!1/6)$ would indicate the breakdown of the procedure of expanding
values in terms of $(\xi\!-\!1/6)$. In such case numerical analysis
would be required.

\section{Conclusion}
\label{sec:conc}

In this paper, we have investigated the scale dependence of the density
perturbations that are generated through modulated reheating after
conformal inflation. We have written down the spectral index and its
running in terms of the flatness parameters and the inflaton's deviation
from an exact conformal coupling. The result indicates the presence of a
large running, which reflects the fact that the derivatives of the
flatness parameters are not necessarily small. This is a basic feature
of conformal inflation, and it is expected that a large running will
be obtained even if we consider other mechanisms for generating
fluctuations (e.g. curvaton
models~\cite{Enqvist:2001zp,Lyth:2001nq,Moroi:2001ct}). It would be
interesting to 
analyze systematically the differences in perturbations generated through
conformal inflation models and standard minimal ones.  

We also applied our results to the warped brane inflation model, where it
was shown that observables were dominantly determined by the deviation
from the conformal coupling. We have shown that the spectral index and
its running are related by (\ref{relation}) for this model.
Since the running is highly constrained by
the WMAP5 data, a stringent bound on the coupling of the inflaton to
gravity was obtained. Also, comparison with the WMAP5 data suggested a
blue tilt of the spectrum. (However, we note that constraints on the
spectrum can be altered if cosmic strings are produced after brane
inflation, see \cite{Polchinski:2004ia,Bevis:2007gh,Battye:2007si} and
references therein.) 

While our analysis focused on general aspects of modulated reheating
after conformal inflation, in this paper we have not presented the light
modulus responsible for generating fluctuations in a concrete setup. In
the warped brane inflation case, potential candidates for such light
fields are angular positions of the D3($\overline{\mathrm{D}3}$)-branes
sitting in throats with angular isometries, and/or axions associated
with shift symmetries of the K\"ahler potential. They may have
negligible effects on the inflaton dynamics during inflation, but one
can expect such fields to play the role of the moduli fields and
eventually generate density perturbations. For further study, it
is important to come up with an explicit realization of our mechanism
based on fundamental theories. We leave this for future work. 

We have focused on the tilt of the density perturbations, but it would
also be worthwhile to study non-Gaussianities in conformal inflation
models. One of the general lessons of our work is that a nonminimal
coupling with gravity can drastically change the behavior of inflation.
A special feature of conformal inflation is that higher-order time
derivatives of the Hubble parameter have large values. Cosmological
observations are imposing (not necessarily direct but) important
constraints on such features even at the present stage.

\begin{acknowledgments}
 We are grateful to  Issha Kayo, Shunichiro Kinoshita, Lev Kofman, and
 Tadashi Takayanagi for useful discussions. T.K. would also like to
 thank Katsuhiko Sato for his continuous support. The work of S.M. was
 supported in part by MEXT through a Grant-in-Aid for Young Scientists
 (B) No.~17740134, and  by JSPS through a Grant-in-Aid for Creative
 Scientific Research No.~19GS0219 and through a Grant-in-Aid for
 Scientific Research (B) No.~19340054. This work was supported by World
 Premier International Research Center Initiative (WPI Initiative),
 MEXT, Japan. This work was also supported in part by the Global COE
 Program ``the Physical Sciences Frontier,'' MEXT, Japan. 
\end{acknowledgments}

\appendix

\section{Detailed Discussions of Conformal Inflation}
\label{sec:c}

We give detailed arguments on conformal inflation. Throughout this
appendix, we consider the case of exact conformal coupling $\xi = 1/6$. 

\subsection{Selection Criterion for $c$}
\label{app:A1}

Let us introduce a dimensionless parameter~$\beta$ 
\begin{equation}
 c H \pi =  -V' (1+\beta) 
\end{equation}
which measures the validity of the approximation~(\ref{333}) during
conformal inflation. (One may wonder why we have used $\alpha$ as
defined in (\ref{alpha}) in the body of this paper. This is due to the
sharp breakdown of the approximation~(\ref{333}) when there is a
deviation from $\xi = 1/6$, in which case $\beta$ can become
significantly large.) 
Then by following the same procedure as for $\alpha$, we can derive the
evolution equation of $\beta$,
\begin{equation}
 \frac{\dot{\beta}}{H} = \mathcal{O}(\epsilon^{1/2},\, \eta_c)  
 -\beta\left\{ c - \frac{3 \eta}{c} + \mathcal{O}(\epsilon^{1/2},\,
      \eta_c)\right\} + \mathcal{O} (\beta^2). \label{beta-dot}
\end{equation}
This shows that the condition
\begin{equation}
 c - \frac{3 \eta}{c} > 0 \label{A3}
\end{equation}
is required for $\beta$ to decay and for (\ref{333}) to be an
inflationary attractor. (\ref{beta-dot}) also indicates that conformal 
inflation is unstable for small $|c|$. 

When $\eta$ is negligible, the inflationary condition $\eta_c/c^2
\approx 0$ gives
\begin{equation}
 c = 2-\zeta
\end{equation}
where we have defined $\zeta \equiv \phi V'' /V'$. Then the
condition~(\ref{A3}) demands 
\begin{equation}
 \zeta < 2  
\end{equation}
in order for the existence of the attractor. 

On the other hand, when $\eta$ is non-negligible, $\eta_c/c^2 \approx 0$
sets the constant~$c$ to either of 
\begin{equation}
 c_{\pm} = \frac{1}{2} 
 \left\{ 2-\zeta  \pm \sqrt{(2-\zeta)^2 - 12\eta}\right\}.
\end{equation}
Considering the condition~(\ref{A3}), one can easily show
\begin{equation}
  c_{\pm} - \frac{3 \eta}{c_{\pm}} = \pm \sqrt{(2-\zeta)^2-12
   \eta}. \label{A5}
\end{equation}
Thus we conclude that (\ref{333}) with the larger solution~$c_+$ is the
attractor of conformal inflation.

\subsection{(Non)Relation Between the Inflationary Attractor and Predictions
  of Cosmological Observables}
\label{app:A2}

We have seen in the previous subsection that once $|\beta |$ becomes
small, (\ref{333}) continues to hold as long as the condition~(\ref{A3})
is satisfied. However, in actual cases, the nonlinear terms in 
(\ref{beta-dot}) may initially prevent $|\beta|$ from dropping down to a
value smaller than unity, and it may take some time for the inflaton to fall
into the attractor. Here, we point out that whether or not the inflaton
trajectory approaches the attractor~(\ref{333}) is not directly related
to the validness of our results~(\ref{result1}) and (\ref{result2}) on
cosmological observables. 

This can easily be understood by recalling how $\alpha$ (or $\beta$)
showed up in (\ref{dotH-alpha}). Taking into account $\alpha = 
\sqrt{2} e \beta$, one can obtain the full expression for the spectral
index in the exactly conformally coupled case
\begin{align}
 n_s - 1 &=
 \frac{4(1-\tilde{\epsilon})-4\sqrt{1+\frac{12}{c^2}(1+\beta)^2
	 \epsilon}} 
{3-2\tilde{\epsilon}-\sqrt{1+\frac{12}{c^2} (1+\beta)^2 \epsilon}} \\
 &= 
 -2\tilde{\epsilon} 
 - \left(  \frac{12}{c^2}(1+\beta)^2+\kappa^2 \right) \epsilon +
 \mathcal{O}(\epsilon^{3/2}) ,
 \label{ns-beta}
\end{align}
where in the last line we have performed the flatness parameter expansion 
without imposing any assumptions on the value of $\beta$. This implies
that even if the inflaton 
trajectory has not settled down to the attractor~(\ref{333})
(i.e. $|\beta| \gtrsim 1$), if 
$\epsilon$ is small enough to satisfy $|\tilde{\epsilon}| \gg \beta^2
\epsilon$ (i.e. $\phi^2/M_p^2 \gg \beta^4 \epsilon$), then the
$\epsilon$ term containing $\beta$ is negligible and our
result~(\ref{result1}) is valid. Similar arguments can also be made for
the running.


\end{document}